\documentclass[traditabstract]{aa}  
\usepackage{graphicx}
\usepackage{lscape}
\usepackage{txfonts}
\usepackage{amssymb}
\usepackage{natbib} 
\bibpunct{(}{)}{;}{a}{}{,} 

\def\aap{A\&A}
\def\apjl{ApJL}
\def\apjs{ApJS}

\def\HI{\mbox{H\,{\sc i}}~}
\def\HInoSp{\mbox{H\,{\sc i}}}

\begin{document}
   \title{The JCMT Nearby Galaxies Legacy Survey VI: The distribution of gas and star formation in M81}
   \titlerunning{The distribution of gas and star formation in M81}
   
   \author{J.~R.~S\'{a}nchez-Gallego\inst{\ref{IAC},\ref{DepAstro}}\thanks{E-mail: jrsg@iac.es}
          \and
          J.~H.~Knapen\inst{\ref{IAC},\ref{DepAstro}}
          \and
		  J.~S.~Heiner\inst{\ref{LavalUni}}
		  \and
          C.~D.~Wilson\inst{\ref{McMaster}}
          \and
          B.~E.~Warren\inst{\ref{McMaster},\ref{CRAR}}
          \and
          R.~J.~Allen\inst{\ref{STScI}}
          \and       
          M.~Azimlu\inst{\ref{WestUni}}
          \and
          P.~Barmby\inst{\ref{WestUni}}
          \and
          G.~J.~Bendo\inst{\ref{ImpColl}}
          \and
          S.~Comer\'{o}n\inst{\ref{KASI}}
		  \and
          F.~P.~Israel\inst{\ref{LeidenUni}}
		  \and
          S.~Serjeant\inst{\ref{OpenUni}}
          \and
          R.~P.~J.~Tilanus\inst{\ref{JAC},\ref{NOSC}}
          \and
          C.~Vlahakis\inst{\ref{LeidenUni}}
          \and
          P.~van~der~Werf\inst{\ref{LeidenUni}}
          }
   \authorrunning{J.~R.~S\'{a}nchez-Gallego et al.}
   \institute{Instituto de Astrof\'{i}sica de Canarias, E-38205 La Laguna, Tenerife, Spain \label{IAC}
            \and
             Departamento de Astrof\'{i}sica, Universidad de La Laguna, E-38200 La Laguna, Tenerife, Spain\label{DepAstro}
             \and
             Department of Physics, Laval University, Qu\'{e}bec City, Qu\'{e}bec, G1V 0A6, Canada\label{LavalUni}
             \and
             Department of Physics \& Astronomy, McMaster University, Hamilton, Ontario, L8S 4M1, Canada\label{McMaster}
             \and
             International Centre for Radio Astronomy Research, M468, University of Western Australia, Crawley, WA, 6009 Australia\label{CRAR}
             \and
             Space Telescope Science Institute, Baltimore, MD 21218, USA.\label{STScI}
             \and
             Physics and Astronomy Department, University of Western Ontario, 1151 Richmond Street, London, Ontario, N6A 3K7, Canada\label{WestUni}
             \and
             Astrophysics Group, Imperial College, Blackett Laboratory, Prince Consort Road, London SW7 2AZ, UK\label{ImpColl}
             \and
             Korea Astronomy and Space Science Institute, 61-1 Hwaam-dong, Yuseong-gu, Daejeon 305-348, Republic of Korea\label{KASI}
             \and
             Sterrewacht Leiden, Leiden University, P.O. Box 9513, 2300 RA Leiden, The Netherlands\label{LeidenUni}
             \and
             Department of Physics \& Astronomy, The Open University, Milton Keynes Mk7 6AA, UK\label{OpenUni}
             \and
             Joint Astronomy Centre, 660 N. A’ohoku Pl., University Park, Hilo, HI 96720, USA\label{JAC}
             \and
             Netherlands Organisation for Scientific Research, Laan van Nieuw Oost-Indie 300, NL-2509 AC The Hague, the Netherlands\label{NOSC}             
             }

   \date{Received May 04, 2010; accepted November 12, 2010}
 
  \abstract{
  We present the first complete $^{12}$CO $J=3-2$ map of M81, observed as part of the Nearby Galaxies Legacy Survey being carried out at the James Clerk Maxwell Telescope. We detect nine regions of significant CO emission located at different positions within the spiral arms, and confirm that the global CO emission in the galaxy is low. We combine these data with a new H$\alpha$ map obtained using the Isaac Newton Telescope and archival \HInoSp, 24$\,\mu\rm m$, and FUV images to uncover a correlation between the molecular gas and star forming regions in M81. For the nine regions detected in CO $J=3-2$, we combine our CO $J=3-2$ data with existing CO $J=1-0$ data to calculate line ratios. We find that the ratio $J=(3-2)/(1-0)$ is in agreement with the range of typical values found in the literature $(0.2-0.8)$. Making reasonable assumptions, this allows us to constrain the hydrogen density to the range $(10^3-10^4)\,\rm cm^{-3}$. We also estimated the amount of hydrogen produced in photo-dissociation regions near the locations where CO $J=3-2$ was detected.}


   \maketitle

%
%
%
%

\section{Introduction}

\defcitealias{Wilson2009}{Paper I}

The details of the interplay between the gas content of galaxies and the processes triggering star formation form one of the main open questions in astronomy (the various reviews and panel discussion transcripts in Knapen et al. 2008 give a broad overview of the topic). Although it is known that star formation occurs mainly in the interior of giant molecular clouds (GMCs), the role played by the different phases of the gas is still poorly understood. For example, \citet{Kennicutt2007} demonstrated that molecular gas traces star formation significantly better than atomic gas. However, the lack of a permanent dipole moment makes the direct detection of molecular hydrogen difficult. This forces us to use tracers such as CO to calculate the amount and distribution of molecular gas. The conversion factor from CO to H$_2$ (the so-called $X\mathrm{_{CO-H_2}}$ factor) has a value that must be considered carefully and the relation of which with the morphological type, metallicity, and other parameters of galaxies is still disputed \citep[e.g.,][]{Maloney1988,Wilson1995,Wada2005,Bolatto2008,Pineda2008,Zhu2009}.

On the other hand, the idea that star formation is a one-way process from gas to stars is an exceedingly simplistic scenario. It has been argued (see, e.g., \citealt{Allen1985,Allen1997,Smith2000}; \citealt{Heiner2008a,Heiner2008b}) that at least part of the atomic gas observed could be produced by photodissociation of molecular hydrogen. In regions with strong star formation, the young, hot stars produce vast amounts of UV ionizing photons that can dissociate an important part of the molecular gas in the host GMC. Therefore, the ability to trace the distribution of molecular gas with good spatial resolution is of vital importance if we wish to understand the behaviour of all the factors involved in the star formation process.

This paper intends to shed further light on these issues by presenting the first complete $^{12}$CO $J=3-2$ map covering the inner part of the optical disk of M81, obtained as part of the Nearby Galaxies Legacy Survey (NGLS) on the James Clerk Maxwell Telescope (JCMT). Several authors (see, for example, \citealt{Wilson2009} and \citealt{Iono2009}) show that $^{12}$CO $J=3-2$, being more energetic and of higher critical density ($n_{\rm c}\sim2.5\times10^4\,\rm cm^{-3}$) than  $^{12}$CO $J=1-0$ ($n_{\rm c}\sim700\,\rm cm^{-3}$), is a better tracer of the warm, dense gas involved in star formation. Thus, this new CO map, in conjunction with new high-resolution H$\alpha$ data from the Isaac Newton Telescope (INT) and public FUV, \HInoSp, and IR images, offers the possibility to study in detail the interplay between atomic and warm molecular gas and their relation with star formation and photodissociation processes.

The JCMT is currently hosting a series of Legacy Surveys, including the NGLS that aims to observe 155 nearby galaxies (within 25\,Mpc). The sample is selected to cover a wide spectrum of morphologies and gas content and overlaps partially with the \emph{Spitzer} Infrared Nearby Galaxies Survey (SINGS; \citealt{Kennicutt2003}) sample. The project entails two different phases: a first set of observations to obtain $^{12}$CO $J=3-2$ observations for the whole sample has already been completed, while the second phase, involving SCUBA-2 continuum observations at 450 and $850\,\mathrm{\mu m}$, will start imminently. Further information about the survey and its science goals can be found in \citet{Wilson2009}.

M81 is a nearby (3.63\,Mpc, \citealp{Freedman1994}) nearly face-on SA(s)a\underline{b} \citep{Buta2007} spiral galaxy. The spatial scale is 17.5\,pc per arcsec, which allows us a detailed study of its GMCs (which have typical sizes between 30 and 50\,pc; see, for example, \citealt{Liszt1981}). Some basic properties of M81 can be found in Table \ref{tab:M81Props}. Although M81 has been extensively studied at multiple wavelengths, just one complete CO map has been released to date. \citet{Brouillet1988} observed M81 in the $J=1-0$ transition, finding weak CO emission across the galaxy. Subsequent observations in the same transition (\citealp{Brouillet1991}; \citealp{Sage1991}; \citealp{Sage1993}; \citealp{Sakamoto2001}; \citealp{Helfer2003}; \citealp{Knapen2006}; \citealp{Casasola2007}) have focused on small, well-defined regions, and confirm the apparent lack of strong molecular emission in M81.

We begin by introducing (in Section \ref{sec:Observ}) the observations and data processing employed in this study. A description of the main results obtained, including a detailed comparison of the selected regions through different wavelengths, is given in Section \ref{sec:Detailed} and a comparison with the photo-dissociation regions (PDRs) method of \citet{Heiner2008a} in Section \ref{sec:PDRs}. We calculate line ratios and gas masses in Section \ref{sec:LineRatiosGasMass}. Conclusions are presented in Section \ref{sec:Conclusions}.

\begin{table}
 \caption{Summary of M81 properties.}
 \label{tab:M81Props}
\centering
 \begin{tabular}{ll}
 \hline
  Parameter & Value$^a$ \\
 \hline
	Object name & M81, NGC\,3031, UGC\,05318 \\
	R.A. (J2000.0) & 9:55:33.5 \\
	Dec. (J2000.0) & +69:04:00 \\
	$B$ magnitude & 7.89 \\
	Classification & SA(s)a\underline{b}$^b$\\
	Heliocentric velocity ($\mathrm{km\,s^{-1}}$) & $−34\pm4$ \\
	Distance (Mpc) & 3.64$\pm0.34^c$ \\
	$D_{25}$ (arcmin) & $26.9\times14.1^d$ \\
	$\rm 12+log(O/H)$ & 8.41$^e$\\
	Scale (pc/arcsec) & 17.5$^c$ \\
	\hline
\end{tabular}
\tablefoot{
   \tablefoottext{a}{Data from \citet{de-Vaucouleurs1991} except where indicated otherwise}
   \tablefoottext{b}{Data from \citet{Buta2007}}
   \tablefoottext{c}{\citet{Freedman1994}}
   \tablefoottext{d}{Major and minor axis}
   \tablefoottext{e}{\citet{Moustakas2006}}
   }
\end{table}

%
%
%
%

\section{Observations and data reduction}
\label{sec:Observ}
\subsection{CO $J$=3-2 map}
\label{sec:COMap}

\begin{landscape}
  \begin{figure}
      \centering
      \vbox to10mm {}
      \includegraphics[width=220mm]{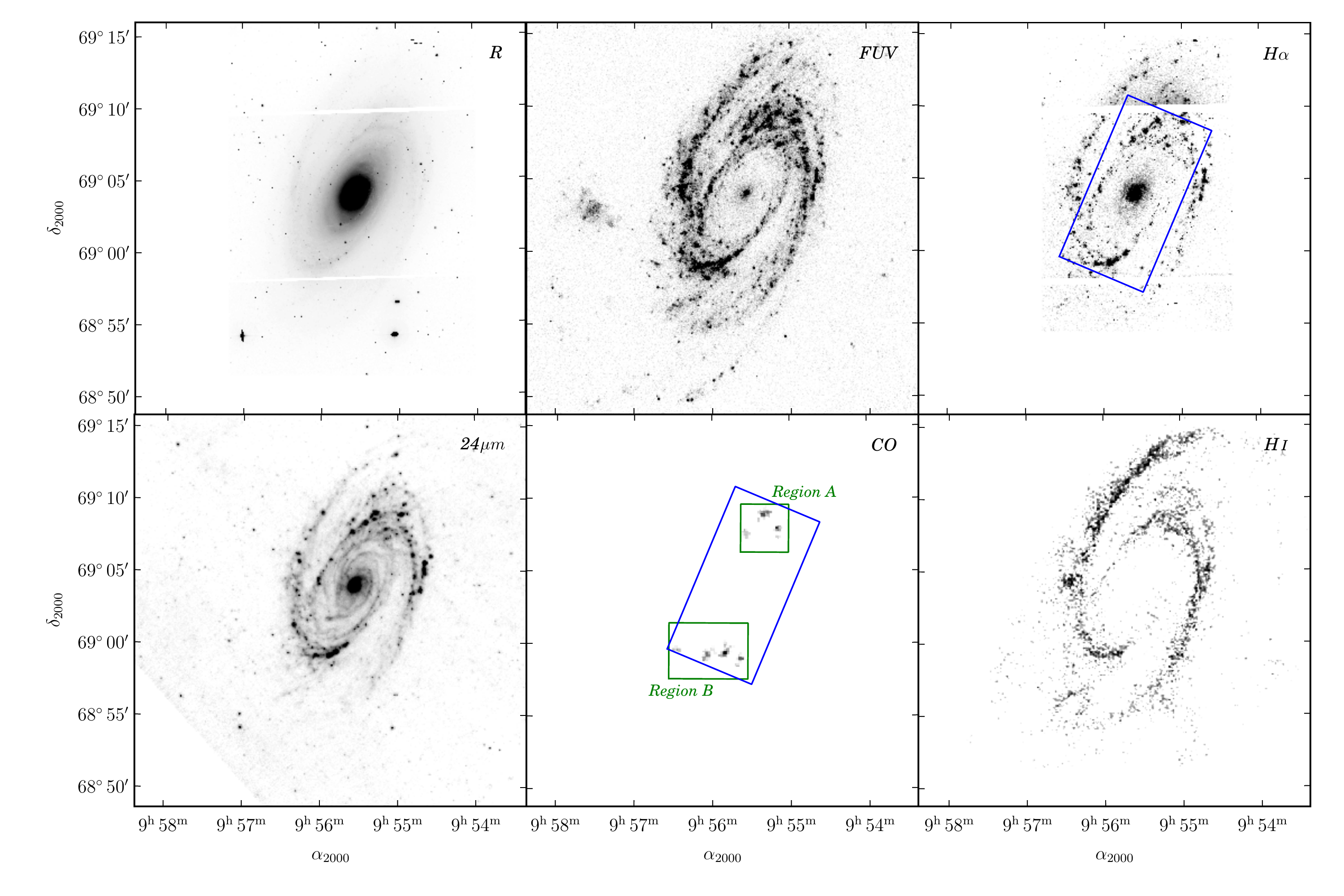}
      \caption{Panels showing, at the same scale, our new CO $J=3-2$, R-band, and continuum-subtracted H$\alpha$ maps together with ancillary data introduced in section \ref{sec:Ancillary}. All the maps are displayed at their original resolution. The cut levels have been arbitrarily set to enhance the main features at each wavelength. The blue solid rectangle represents the CO $J=3-2$ region mapped. For comparison purposes, the same area is also overlaid on the H$\bf\alpha$ map. The regions where the CO emission is not reliable (see Section \ref{sec:Observ}) are not shown. Green lines delimit the areas covered by the regions displayed in Figures \ref{fig:RegionA} and \ref{fig:RegionB}, where all our CO detections are located. The white spaces shown in the H$\alpha$ map are due to the separation between the chips in the CCD array.}
      \label{fig:Maps}
  \end{figure}
\end{landscape}

The $^{12}$CO $J=3-2$ data presented here were obtained as part of the JCMT NGLS. The observations were taken in raster map scanning mode during three nights in February 2008 using the 16 receptors heterodyne array receiver (HARP) with the auto-correlation spectrometer imaging system \citep{Smith2003} as backend \citep{Wilson2009,Warren2010} The spectrometer was configured to achieve a bandwidth of 1\,GHz with a resolution of $0.488\,\rm MHz$ ($0.43\,\mathrm{km\,s^{-1}}$ for the studied $^{12}$CO $J=3-2$ transition). The JCMT beam at this wavelength is 14.5\,arcsec FWHM and the signal was sampled using 7.276\,arcsec pixels to satisfy the Nyquist criterion. Given the 17.5\,arcsec spatial scale, this resolution allows us to resolve features separated $\sim250\,$pc from each other. The observed region for M81 covers a rectangular area of 13.2$\times$7.7\,arcmin, including the brightest regions in H$\alpha$, IR, and FUV (see Figure \ref{fig:Maps}). Following the criteria used for the whole NGLS \citep{Wilson2009}, the galaxy was sampled until the target noise parameters ($T\rm_A^*=19\,mK$ at a resolution of $\rm20\,km\,s^{-1}$) were reached.

The data reduction process is explained in detail in \citet{Wilson2009} and in Appendix A of \citet{Warren2010}, so only a brief summary of the given steps is presented here. After flagging bad baselines and spikes, the individual scans were combined using a $\textnormal{sinc}(\pi x)\textnormal{sinc}(k\pi x)$ weighting function. The final cube was trimmed to remove the edges of the scans (approximately 60\,arcsec) where coverage is incomplete. To remove baselines, line regions were masked out (applying different methods for narrow and broad lines and combining the two resulting masks) and third order polynomials were fitted to the continuum. These baselines were then subtracted from the original spectra. 

To compile the integrated CO map, we used a two-step approach. In the first step, we used the \textsc{clumpfind} algorithm \citep{Williams1994} to select regions with emission above three times the RMS noise level (3$\bf\sigma$), based on the contour levels defined by \textsc{clumpfind} in both the physical and spectral dimensions and on the number of channels in the data cube and the noise in each channel. This RMS value ensures, based on our experience, that all the potential features present in the CO map are found. We then applied these masks to produce moment maps from the original cube \citep{Warren2010}. The data values were converted to main beam temperature by dividing by $\eta_{\rm MB}=0.6$.

\begin{figure*}
    \includegraphics[width=170mm]{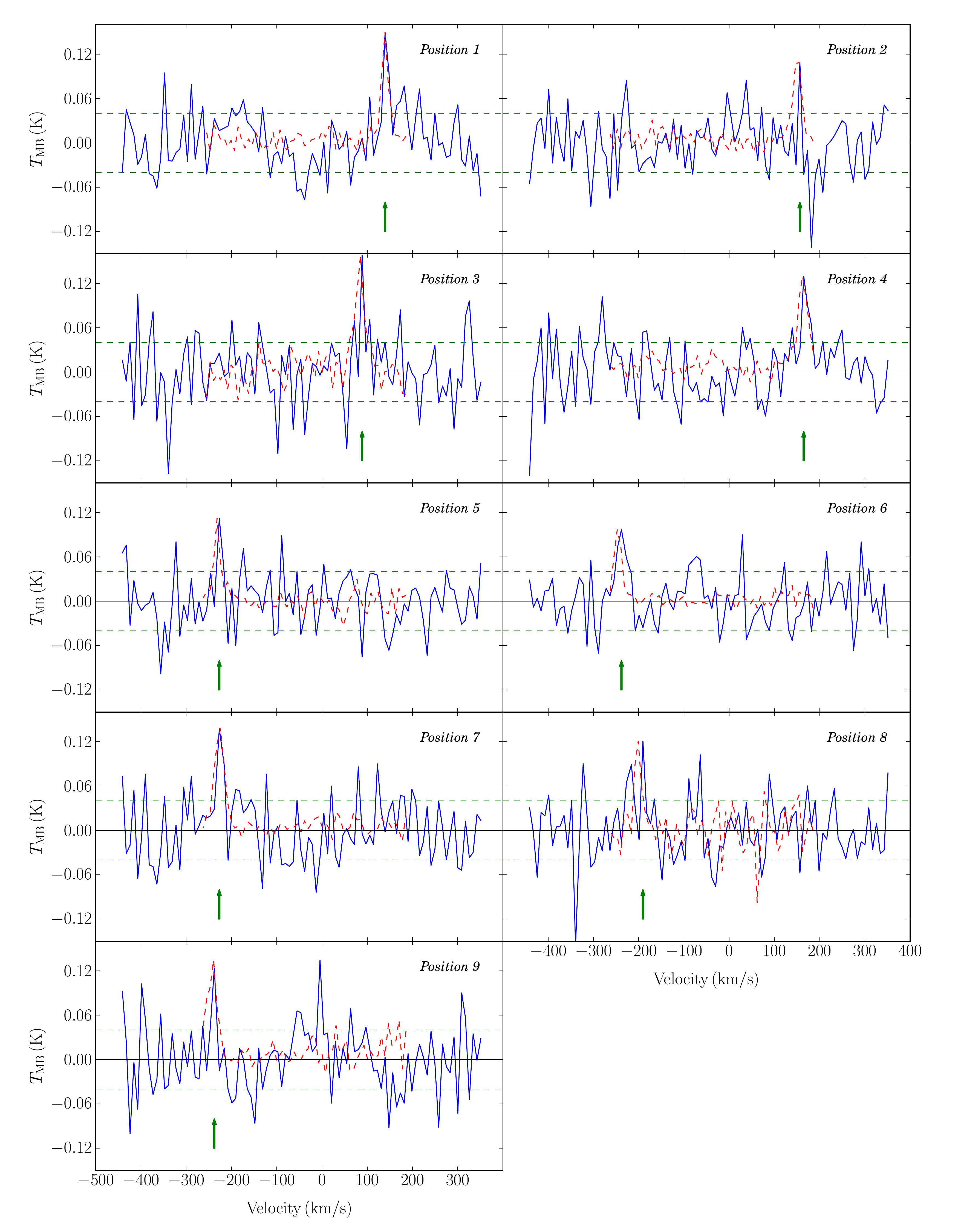}
    \caption{CO $J=3-2$ spectra (blue solid) obtained from the peak pixel of each position where we detected CO, compared to the \HI spectra (red dashed) at those positions. In each case the spectra have been smoothed to $\Delta v=10\,\rm km/s$ and the \HI profile has been scaled to fit the CO peak (pointed out with an arrow at each position). No spatial smoothing was applied. The green dashed lines mark the $\sigma\sim\pm0.04\,\rm K$ average level (see Table \ref{tab:SpectraStats}). The positions labeled are the same as in Figures \ref{fig:RegionA} and \ref{fig:RegionB}.}
    \label{fig:Spectra}
\end{figure*}

In the second step, we imposed a different and more restrictive criterion to each region in the moment map resulting from the first step, aiming to reach a detailed confirmation of every individual feature in the \textsc{clumpfind} CO map. We considered a region to be a significant detection only if the peak value of the integrated spectrum of the region was above 2.5$\bf\sigma$, and if its velocity agreed with that expected from an \HI velocity field.  For the latter, we compared the CO spectra at each position with an \HI profile from THINGS \citep{Walter2008} \HI data cube (see Section \ref{sec:Ancillary} for further details about these data). For each CO position, we compared the peak of CO emission with that at the equivalent position in the \HI map, after smoothing both spectra to a channel velocity of 10$\,\rm km\,s^{-1}$. Figure \ref{fig:Spectra} shows the CO and \HI spectra for each detected feature. The agreement between the spectra, in terms of peak velocity and line width, is good in almost every case. The average separation between the CO and \HI peaks is $ 4.4\pm3.3\,\rm  km\,s^{-1}$.  One spurious feature was rejected because its line velocity was in complete disagreement with that of neighbouring CO regions and with the expected velocity as derived from published \HI data (see below).

As a result of this second step, a total of four features were manually rejected as their line emission peak was confirmed to be significantly below 2.5$\sigma$, or their line velocities or widths were different from the expected values from CO $J=1-0$ and \HI data (\citealt{Helfer2003,Knapen2006,de-Blok2008}). After these verifications, we confirm the detection of nine peaks of 2.5$\sigma$ or more in M81. The final total intensity (moment 0) map for M81 is displayed, along with the other maps used in this paper, in Figure \ref{fig:Maps}.

To quantify the uncertainty in our profiles, we estimated the noise level (by masking the CO line and calculating the standard deviation of the remaining spectrum) and compared it with the value of the peak. The results, shown in Table \ref{tab:SpectraStats}, confirm that all our detections are around 3$\sigma$ or better. The noise levels calculated are in good agreement with the global values measured in other galaxies of the NGLS (\citealp{Wilson2009}; \citealp{Warren2010}; \citealp{Bendo2010}). Table \ref{tab:SpectraStats} also includes the integrated flux $(\int T_{\rm MB}{\rm dv})$ and the FWHM for each spectral line derived from a Gaussian fit.

\begin{table*}
   \caption{CO spectral analysis.}
   \label{tab:SpectraStats}
   \centering
   \begin{tabular}{ccccccccc}
       \hline
       Position & RA$^a$ & Dec$^a$ & Peak velocity & Peak value $(T_{\rm MB})$ & $\sigma\rm^b$ & Peak-$\sigma$ ratio & FWHM$\rm^c$ & $\int T\rm_{MB}dv^c$ \\
		  &        &         & $\rm km\,s^{-1}$ & $\rm K$ & $\rm K$ &  & $\rm km\,s^{-1}$
		  &  $\rm K\,km\,s^{-1}$ \\
       \hline
       1  &  9:55:19.0  &  +69:08:47  &   139.5  &  0.15  &  0.041  &  3.7  &  15    &  2.33  \\
       2  &  9:55:15.5  &  +69:08:55  &   156.4  &  0.11  &  0.039  &  2.8  &  6.5   &  0.76  \\
       3  &  9:55:33.0  &  +69:07:28  &    88.7  &  0.16  &  0.047  &  3.4  &  8.5   &  1.46  \\
       4  &  9:55:08.5  &  +69:07:52  &   164.9  &  0.13  &  0.043  &  3.0  &  17    &  2.25  \\
       5  &  9:56:25.0  &  +68:59:33  &  -227.0  &  0.11  &  0.038  &  3.0  &  13    &  1.55  \\
       6  &  9:56:04.0  &  +68:59:01  &  -238.0  &  0.10  &  0.035  &  2.8  &  28    &  2.81  \\
       7  &  9:55:48.0  &  +68:59:20  &  -227.0  &  0.14  &  0.044  &  3.1  &  19    &  2.73  \\
       8  &  9:55:37.5  &  +68:59:04  &  -190.6  &  0.12  &  0.042  &  2.9  &  8.2   &  1.09  \\
       9  &  9:56:03.0  &  +68:59:15  &  -238.0  &  0.12  &  0.046  &  2.7  &  12    &  1.56  \\
       \hline
   \end{tabular}
   \tablefoot{
       \tablefoottext{\rm a}{Right ascension and declination of the peak pixel (J2000)}
       \tablefoottext{\rm b}{Calculated after masking $\pm50\,\rm km\,s^{-1}$ around the line. Average $\sigma$ is 0.042 at $10\rm\,km\,s^{-1}$ resolution.}
       \tablefoottext{c}{Obtained using Gaussian fits.}
   }
\end{table*}

\subsection{\emph{R} and H$\mathbf{\alpha}$ images}

We present new \emph{R} and H$\alpha$ images acquired using the Isaac Newton Telescope in the course of a programme that aims to obtain H$\alpha$ continuum-subtracted images for all the JCMT NGLS galaxies. The Wide Field Camera (WFC) was used to obtain three H$\alpha$ exposures of $480\,\rm s$ each and two \emph{R}-band exposures of $120\,\rm s$ each. The WFC consists of four 2048$\times$4100\,pixel CCDs with 1\,arcmin separation between the chips and a pixel scale of 0.33\,arcsec per pixel. The final mosaic covers an area of approximately $33\times21$\,arcmin in right ascension and declination respectively. Although maps of similar resolution have been released in the past (for example, as part of the ancillary data for the \emph{Spitzer} infrared nearby galaxies survey, SINGS, \citealt{Kennicutt2003}), our new map covers a larger area and is deeper than other maps of M81 published to date. The spatial resolution of the images is $\sim1.2$\,arcsec FWHM, which will allow us to detect features of the order of the typical size of GMCs. 

The reduction was carried out using IRAF\footnote{http://iraf.noao.edu/} following the prescriptions given in \citet{Knapen2004} adapted for the WFC. After the bias and flat field correction, the images were normalized by dividing by the exposure time and then aligned and combined using a median rejection algorithm in order to remove cosmic rays from the final image. The background level was calculated by taking an area with little emission, rejecting those pixels with values 3$\sigma$ above or below the mean, and iterating until convergence was achieved. This value was then subtracted from the image. Some defects caused by scratches on the CCDs were corrected, replacing the affected pixels with a linear interpolation of the nearest good pixels.

At this point, we aligned the final H$\alpha$ and \emph{R}-band background-subtracted images. To subtract the continuum emission from the H$\alpha$ map, we plotted the intensity, in counts per second, of each pixel in the H$\alpha$ image versus the intensity of the same pixel in the \emph{R}-band image. Three percent of the brightest pixels were manually rejected to avoid the contribution of foreground stars. The resulting plot should display a straight line in the absence of H$\alpha$ emission, and the slope of the line fitted to the upper limit of the distribution provides the scaling factor between both images (see Figure 1 in \citealt{Knapen2004}). Using this scale factor, the \emph{R}-band continuum image was scaled and subtracted from the H$\alpha$ image. The result is shown in Figure \ref{fig:Maps}.

\subsection{Ancillary data}
\label{sec:Ancillary}

We used the VLA \HI data cube obtained by B. Hine and A. Rots \citep{Hine1984}. This final column density map is the same as that used in \citet{Allen1997} and the details concerning the calculation of the moment map can be found there. The spatial resolution of the image is 9\,arcsec. A new deeper \HI moment map of M81, based on archival VLA data, was released by the THINGS team \citep{Walter2008}. The resolution of these new data is slightly better: 7.58$\times$7.45\,arcsec. Unfortunately, this map is not continuum subtracted and cannot be used for our comparison with other wavelengths. However, as described in Section \ref{sec:COMap}, we use the THINGS data to compare the position and width of the spectral lines found in the CO map. The presence of continuum emission is not a problem for this use.

The FUV image, obtained from the \emph{GALEX} public database\footnote{http://www.galex.caltech.edu/index.html}, was acquired as part of the \emph{GALEX} nearby galaxy survey (NGS; \citealt{Gil-de-Paz2004}). The spectral range observed is $1344-1786\,\rm\AA$ with an effective wavelength of 1528$\,\rm\AA$ \citep{Morrissey2007}. The FOV is 1.28\,degrees with a FWHM of 4\,arcsec. The data can be converted into UV flux ($\mathrm{erg\,cm^{-2}\,s^{-1}\,\rm\AA^{-1}\,arcsec^{-2}}$) by applying a conversion factor of $6.22\cdot 10^{-16}$ \citep{Morrissey2007}.

We used the 24\,micron image from SINGS to trace dust emission associated with recent star formation activity (\citealt{Calzetti2005,Calzetti2007,Prescott2007}).  The data were taken using the Multiband Imaging Photometer for Spitzer (MIPS; \citealt{Rieke2004}) in scan map mode.  The final image is $32.5\times55$\,arcmin. The data have a FWHM of 6\,arcsec (Spitzer Observers Manual), and the flux calibration accuracy is 4\% \citep{Engelbracht2007}. The survey website\footnote{http://sings.stsci.edu/} gives further details on the data processing.

For the H$\alpha$, \emph{R}-band, FUV and 24$\rm\,\mu m$ data, the astrometry of the images was checked by comparing the positions of foreground stars with the Guide Star Catalogue \citep{Lasker1990}. It was found that the astrometry of all the images was precise to within 1\,arcsec, which is sufficient for the purposes of this work.

\subsection{CO $J=1-0$ and $J=2-1$ maps}
\label{sec:COBrouillet}

Although CO $\rm J=1-0$ observations of the inner regions \citep{Sakamoto2001, Casasola2007} and the western arm of M81 \citep{Knapen2006} have been performed, the only comprehensive CO $\rm J=1-0$ map available for M81 is the one published by \citet{Brouillet1988} (hereafter the Brouillet map) using 12 meter National Radio Astronomy Observatory (NRAO) data. The half power beam width (HPBW) of the map is 60\,arcsec. This map was later improved with 23\,arcsec HPBW IRAM\footnote{Institut de Radioastronomie Millim\'{e}trique} 30m observations in some regions \citep{Brouillet1991}. However, there is no spatial correspondence between their detections and our CO $\rm J=3-2$ emission, thus the IRAM map cannot be used for further comparisons with our data.

\citet{Sakamoto2001}, in the CO $ J=1-0$ line, and \citet{Casasola2007}, in both the $ J=1-0$ and $J=2-1$ lines, probed a small area in the nuclear region of M81 (40 arcsec, \citeauthor{Casasola2007} and 20 arcsec, \citeauthor{Sakamoto2001}). We convolved our data cube to the resolution of their data and found no CO $J=3-2$ emission in those regions. This is no surprise as, at our current RMS of 19$\,\rm mK$ at 20$\,\rm km\,s^{-1}$ resolution, their lines (of typical temperature $\rm \sim50\,mK$, which with a typical CO $\rm J=1-0$ to 3-2 ratio would correspond to $T\rm_{MB}(3-2)\sim25\, mK$) are too weak to allow CO $\rm J=3-2$ detections at a level of 2.5$\sigma$.

\citet{Knapen2006} used the $\rm 45\,m$ radio telescope at the Nobeyama Radio Observatory (NRO) to observe a small region in the western spiral arm of M81. Unfortunately, this region is outside our mapped area.

\subsection{Convolution kernels}
\label{sec:Kernels}

To compare images at different wavelengths it is necessary to smooth all the data to match the image with the lowest resolution, which in this case is the CO $J=3-2$ map (14.5\,arcsec). For the H$\alpha$, \textit{R}-band, and FUV images, we convolved the original data with gaussians of shape $$\psi(r) \propto {\rm e}^{−r^2/2(\sigma_{\rm CO}^2-\sigma^2)},$$ where $\sigma$ is the original resolution of the image and $\sigma_{\rm CO}$ is the resolution of the CO map. The convolution was performed using the task \textsc{gausmooth} in Starlink/\textsc{KAPPA}.

The MIPS 24\,micron PSF cannot be approximated as a Gaussian function because of its strong Airy rings, so we cannot apply the above equation to smooth the data.  Instead, we smoothed the data using a convolution kernel created by following the methods outlined by \citet{Gordon2008} and \citet{Bendo2010} using an empirical 24\,micron PSF originally derived by \citet{Young2009}. The convolution was performed using the task \textsc{convolve} in Starlink/\textsc{KAPPA} \citep{Currie2008}.

%
%
%
%

\section{Detailed analysis of CO emission regions}
\label{sec:Detailed}

As presented in Figure \ref{fig:Maps}, the emission of the star formation tracers (FUV, H$\alpha$, and 24$\,\mathrm{\mu m}$) follows a common pattern with strong centrally peaked emission and well-traced spiral arms. This is also reflected in the \HI map, except for the lack of nuclear emission. All the detected CO $J=3-2$ features are located in two small regions within the spiral arms. The comparison between H$\alpha$ and FUV shows a good agreement. In almost every case, an individual region with enhanced H$\alpha$ emission has a counterpart in the FUV map. The opposite is not always true. This result confirms those presented by \citet{Allen1997}. In this section, we compare in detail the emission of the different tracers on the smallest resolved spatial scales at the location of the CO emission.

\subsection{CO regions}
\label{sec:CORegions}

\begin{landscape}
  \begin{figure}
      \centering
      \vbox to10mm {}
      \includegraphics[width=220mm]{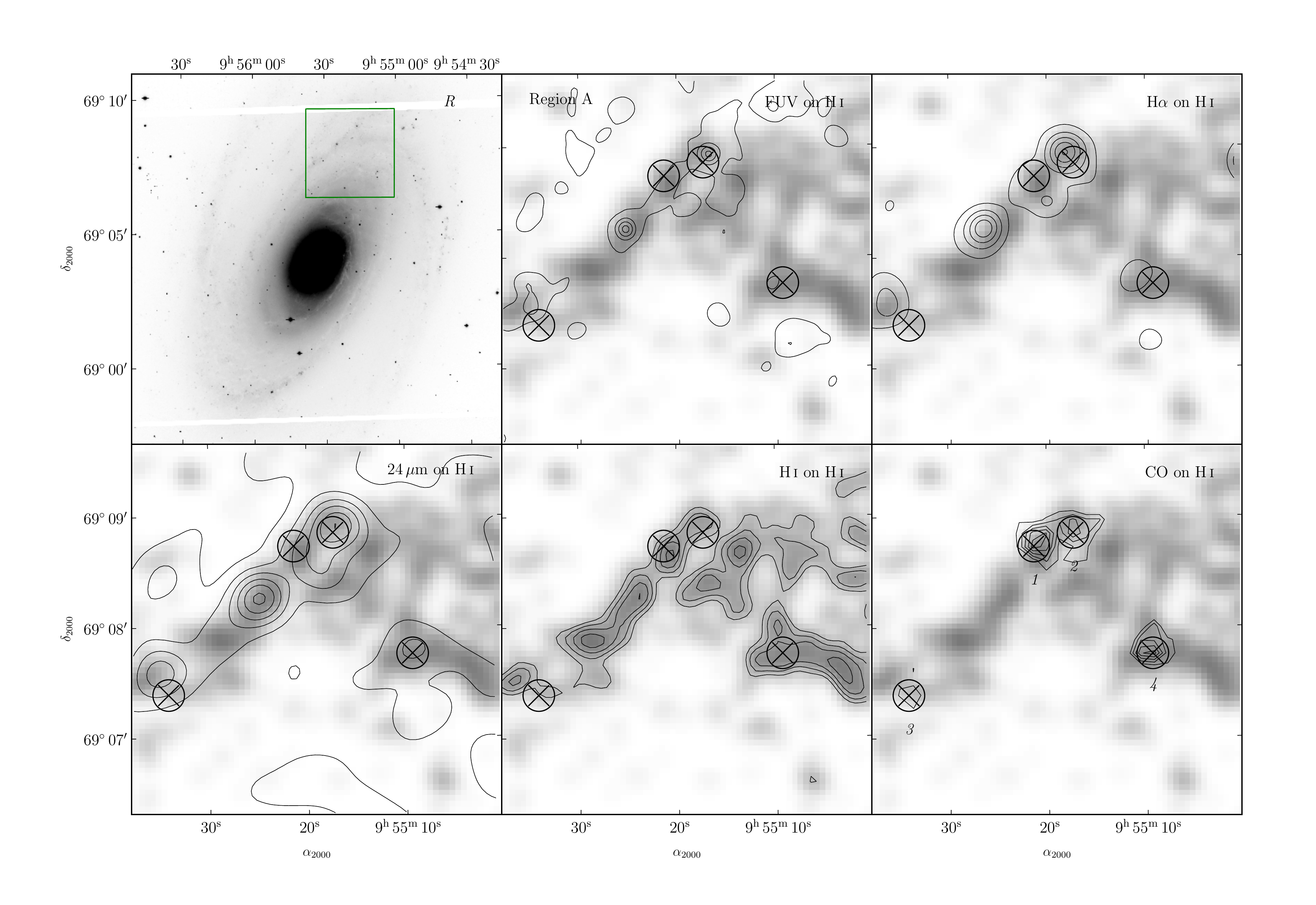}
      \caption{Each panel shows the contours of the indicated wavelength band or image, as labelled, overlaid on the \HI image of region A. Both the \HI map and the contours have been smoothed to match the 14.5\,arcsec resolution of the CO data (see section \ref{sec:Kernels}). FUV levels start at 5$\times10^{-18}\rm\,ergs\,cm^{-2}\,s^{-1}\AA^{-1}\,arcsec^{-2}$ increasing by the same amount for each new contour. \HI contour levels go from 7.68$\times10^{20}$ to 6.53$\times10^{21}$ in steps of 3.84$\times10^{20}$ atoms$\,\rm cm^{-2}$. The CO intensity levels start at 0.4$\,\rm K\,km\,s^{-1}$ ($\sim1\sigma$) and each new level increases 0.2$\,\rm K\,km\,s^{-1}$. H$\alpha$ and 24$\,\mu \rm m$ contour levels are arbitrary. The top left panel shows the footprint of region A plotted on the \emph{R}-band image. Note that the declinations of the FUV and H$\alpha$ panels are as those in the 24$\,\mu\rm m$ panel. The crosses indicate the position of each CO feature, while the circles define a region of 150\,pc around this central value. The identification of the CO positions is given in the bottom right panel and is referred to in Table~\ref{tab:COStats}.}
      \label{fig:RegionA}
  \end{figure}
\end{landscape}

\begin{landscape}
  \begin{figure}
      \centering
      \vbox to10mm {}
      \includegraphics[width=240mm]{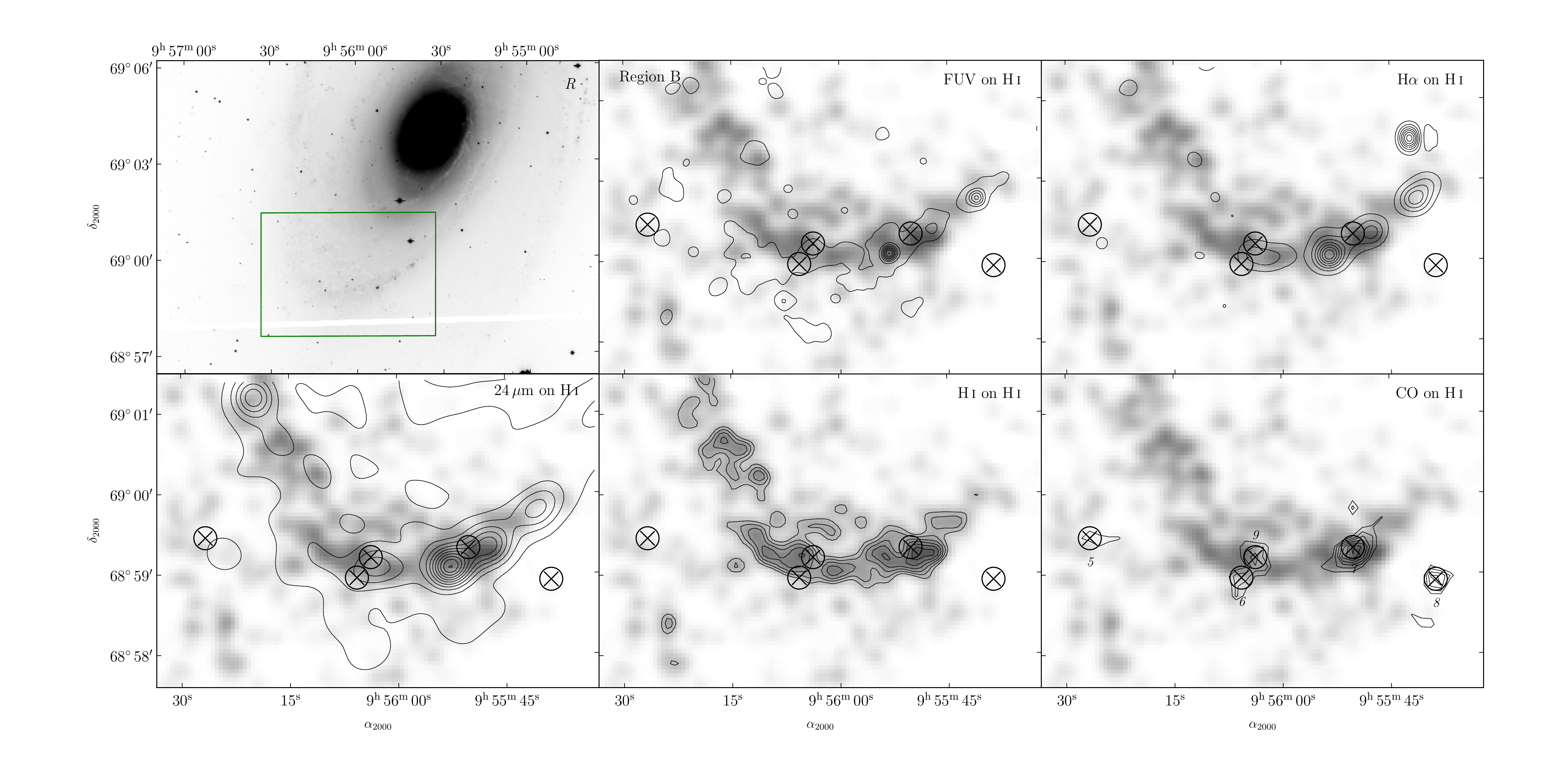}
      \caption{As in Figure~\ref{fig:RegionA}, now for the CO positions in region B.}
      \label{fig:RegionB}
  \end{figure}
\end{landscape}

Our CO emission is located in two regions of the galaxy: one in the northern and another in the southern arm. We call these regions A and B, respectively (see Figure~\ref{fig:Maps}). Figures \ref{fig:RegionA} and \ref{fig:RegionB} show the detected CO features overlaid on the various other maps for each region. Although positions 1 and 2 (region A) are located within the same overall complex, we treat them as two different peaks; this is evident from the CO contour levels (Figure~\ref{fig:RegionA}). The same holds for positions 6 and 9 in region B (Figure~\ref{fig:RegionB}).

A detailed comparison with data at FUV, H$\alpha$, IR, and \HI wavelengths was carried out for each CO feature, with the results shown in Table~\ref{tab:COStats}. Although the typical size of a GMC is in the range of $30-50\,\rm pc$, the accuracy in the determination of position and distances between features is limited by the FWHM of the CO map (14.5\,arcsec, or $\sim$250$\,\rm pc$ in M81). Taking these values into account, we consider two kinds of correlation: the case where a peak of the studied wavelength is found at 50\,pc or less from the peak in the CO feature (marked as P, peak, in Table~\ref{tab:COStats}), and the case when a peak is found in an area of 150\, pc in radius around the peak of CO emission (marked as C, close). This task was performed for each wavelength using the images convolved to the CO resolution.

A peak of emission close to the CO feature is found in all bands for 33\% of the cases, except in H$\alpha$, where the probability is slightly lower (22\%). On the other hand, the probability of finding an emission peak at the exact location of the CO peak is much lower (just one coincidence among all the positions and wavelengths), which indicates that the molecular gas does not occupy the same space as the atomic hydrogen or the ionized gas, on the scales considered.

\begin{table}
   \caption{Statistics of the CO positions.}
   \label{tab:COStats}
   \centering
   \begin{tabular}{cccccc}
       \hline
       Position  &    FUV       &   H$\alpha$  &  $24\,\mu$m &    \HI       \\    
       \hline
          1      &     -        &      C       &      -      &      C       \\     
          2      &     C        &      C       &      C      &      C       \\     
          3      &     -        &      -       &      -      &      C       \\     
          4      &     C        &      -       &      P      &      -       \\     
          5      &     -        &      -       &      -      &      -       \\     
          6      &     -        &      -       &      -      &      -       \\     
          7      &     -        &      -       &      -      &      -       \\     
          8      &     -        &      -       &      -      &      -       \\     
          9      &     C        &      -       &      C      &      -       \\ 
       \hline     
       P (\%)    &     0        &      0       &   $11\pm10$ &      0       \\    
       C (\%)    &   $33\pm16$  &   $22\pm14$  &   $33\pm16$ &   $33\pm16$  \\ 
       \hline 
       \hline
       RP (\%)   &   $3\pm2$    &      0       &   $1\pm1$   &   $1\pm1$    \\    
       RC (\%)   &   $11\pm3$   &   $10\pm3$   &   $4\pm2$   &   $15\pm4$   \\
       \hline
   \end{tabular}
   \tablefoot{For each position and wavelength, the table indicates whether a peak of emission was found at the exact location (P) or if a peak is closer than 150\,pc (C). The two lines below indicate the percentage of detections in each case. The last two lines show the same statistics for the peak (RP) and close (RC) random positions. Note that the close percentages also includes the peak cases. Errors have been calculated using Poisson statistics. }
\end{table}

\subsection{Randomly selected regions}
\label{sec:RandomRegions}

The results presented above seem to indicate that the CO emission is correlated with both \HI and star formation tracers. To place these results in perspective, we used a random number generator to obtain 100 completely random positions within the footprint of regions A and B.

The last two rows of Table~\ref{tab:COStats} present the results of the comparison between these random regions and the emission at the  available wavelengths. We find that the probability of a position lying exactly on a peak is quite low (3\% for FUV, lower for the remaining wavelengths). When a peak is located near the central position (closer than 150\,pc), the numbers do not change significantly: the probabilities increase to 15\% only for \HI and always remain below the percentages obtained for the CO regions.

These values, compared to the significantly higher ones found in the previous section for CO, indicate that a relation between the CO $J=3-2$ emission and the tracers of star formation and atomic gas indeed exists. The correlation is, however, not as strong as in the case of the star formation tracers and \HInoSp, where a correspondence has been found for about 90\% of the peak positions \citep{Allen1997}.

\section{\bfseries Analysis of candidate PDRs}
\label{sec:PDRs}

In principle, it is possible to estimate the total hydrogen volume density of gas near a PDR from the measured UV flux and \HI column density, by making certain assumptions about the physical properties of the PDR. Comparing this estimate with measurements of the molecular gas column density, obtained from CO observations, one can then obtain information about the physics and even internal structure of the PDR, and constrain the CO to $\rm H_2$ conversion factor.

Using the approach outlined in \citet{Heiner2008a}, we investigate the presence of candidate PDRs near the detected CO $J=3-2$ emission. While \citeauthor{Heiner2008a} selected the brightest FUV regions in the galaxy, we used our CO detections as reference points to locate the nearby peaks of UV emission and \HI patches.

The method applied was explained in depth in \citet{Heiner2008a}. However, for the sake of clarity, we include here a short explanation of the main steps. \citet[his equation 6]{Allen2004} derived the following expression for the \HI column density produced by photodissociation near UV sources:

$$\rm
{\mathnormal N}_{{\rm H\,{\mathsc i}}}=\frac{7.8\times 10^{20}}{\delta/\delta_0}\ln\left[1+\frac{106\times {\mathnormal G_0}}{\mathnormal n}\left(\frac{\delta}{\delta_0}\right)^{-1/2}\right]
$$

\noindent{}This relation relies heavily on the dust-to-gas ratio ($\rm\delta/\delta_0$, with $\rm \delta_0$ the local solar neighborhood value), which is derived from metallicity measurements, where $G_0$ is an expression of the incident UV flux at the \HI patch in units of the Habing flux \citep{van-Dishoeck1988, Allen2004} and depends on the distance to the galaxy and the separation between the UV source and the \HI column, $\rho_{\rm{H\,{\mathsc i}}}$. This expression can be inverted to calculate the total density of gas in the GMC, $n\rm(cm^{-3})={\mathnormal n}_{H\,{\mathsc i}}+2{\mathnormal n}_{H_2}$. At the core of the GMC ,we can then expect that $\rm {\mathnormal n}=2{\mathnormal n}_{H_2}$.

The parameters used here are the same as in \citet{Heiner2008a}, except for the extinction [$A_{\rm FUV} = 7.9\times E(B-V) = 0.63$ \citep{Gil-de-Paz2007} with $E(B-V) = 0.080$ \citep{Schlegel1998}] and the calibration of the $\rm \delta/\delta_0$ relation [$\rm\log(\delta/\delta_0)=-0.045\times {\mathnormal R}_{gal} + 0.31$, now using an updated value of solar metallicity \citep{Allende-Prieto2001}]. We used the distance referred to in Table \ref{tab:M81Props}. 

Table \ref{tab:PDRs} shows the values measured for each of our CO detections and the total density of gas calculated for each region. Owing to the lower UV extinction value used in this work, the total density results are approximately half of those that would be obtained with the values used in \citet{Heiner2008a}. The uncertainties were calculated as explained in  Appendix A of \citet{Heiner2008a}. The values obtained are in the range $1-178\,\rm\,cm^{-3}$ for the total hydrogen volume density. The highest value occurs near CO position 2, which was already noted in Table \ref{tab:COStats} to have FUV, H$\alpha$, 24$\,\rm\mu m$, and \HI emission within 150$\,\rm pc$, making it the candidate PDR most likely to be directly comparable to the CO emission at the same location.

\begin{table*}
   \caption{PDR data}
   \label{tab:PDRs}
   \centering
   \begin{tabular}{ccccccccccc}
       \hline
       CO position  &  \multicolumn{2}{c}{FUV coordinate}  &  $R_{\rm gal}$  &  UV flux                                          &  ${\rho_{\rm H\,{\mathsc i}}}$  &  $N_{\rm H\,{\mathsc i}}^{\rm a}$    &  $\rm \delta/\delta_0$  & $G_0$ & $n$             & uncertainty in $n$ \\
                        &  R.A.  &  Dec.                   &    kpc            &  $\rm 10^{-15}\,ergs\,cm^{-2}\,s^{-1}\,\AA^{-1}$  &  pc                                          &  $\rm 10^{21}\,cm^{-2}$  &                             &               &  $\rm cm^{-3}$ & \bf\%                \\   
		\hline

       1$^{b,c}$     & 09:55:16.75 & +69:08:55.62 & 5.6 & 4.11 & 469 & 2.50 & 1.139 &  0.14 & 0.4 & 44  \\
       2$^{b}$       & 09:55:16.75 & +69:08:55.62 & 5.6 & 4.11 & 26  & 2.28 & 1.139 & 47.93 & 178 & 146 \\
       3             & 09:55:35.02 & +69:07:34.69 & 4.7 & 6.99 & 81  & 2.93 & 1.251 &  8.17 & 7   & 69  \\
       4             & 09:55:10.25 & +69:07:46.30 & 4.7 & 0.70 & 182 & 2.63 & 1.260 &  0.16 & 0.2 & 53  \\
       5             & 09:56:24.69 & +68:59:17.12 & 8.0 & 1.59 & 251 & 1.55 & 0.887 &  0.19 & 5   & 31  \\
       6/9$^{d}$     & 09:56:03.79 & +68:59:05.59 & 6.0 & 4.73 & 278 & 4.31 & 1.100 &  0.47 & 0.1 & 68  \\
       7             & 09:55:47.43 & +68:59:23.33 & 5.1 & 5.46 & 95  & 4.26 & 1.205 &  4.61 & 0.6 & 81  \\
       8$^{e}$       & 09:55:37.50 & +68:59:04.00 & 6.0 & 0.15 & 170 & 0.92 & 1.096 &  0.04 & 1.8 & 32  \\
       \hline
   \end{tabular}
   \tablefoot{
       {See Section \ref{sec:PDRs} for an explanation of the parameters in this table}
       \tablefoottext{a}{\HI background subtracted was $\rm 10^{20}\,cm^{-2}$}
       \tablefoottext{b}{CO positions 1 and 2 have one FUV source in the middle}
       \tablefoottext{c}{A weak FUV source is present near position 1 but its flux could not be accurately measured}
       \tablefoottext{d}{One measurement was carried out for CO positions 6 and 9}
       \tablefoottext{e}{Upper limit for a source at the GALEX detection limit ($F\rm_{FUV}=0.15\times10^{-15}\,ergs\,cm^{-2}\,s^{-1}\,\AA^{-1}$).}
       }
\end{table*}

We speculate that the gas clouds traced by the PDRs are fundamentally different from those detected in CO $J=3-2$. The latter are indicative of a dense, hot environment, concentrated in a relatively small volume of space, where the PDR method is sensitive to lower density gas. A molecular cloud like this does not have to contain an OB association. For example, CO position 8 corresponds to a known \mbox{H\,{\sc ii}} region that is too faint to be detected in the GALEX image that we used. It might also correspond to a very dense core of a larger association of GMCs. An important additional clue to \HI being produced in PDRs, in turn produced by OB star clusters, would be the presence of bubbles of \HI, where the atomic gas has been cleared by recent star formation. Comparing the locations of our CO detections with the \HI holes presented in \citet{Bagetakos2010}, we find that the majority of our detections occur on the edge of an \HI hole (the smallest \HI hole detected being $90\rm\,arcsec$ across). Exceptions are positions 4, 7, and 8 where no \HI holes were detected in the immediate vicinity.

The gas densities derived with the PDR method are most likely to yield lower limits --we take the maximum \HI column density, where the higher-density, more-compact PDRs might produce a lower \HI column density because the column is developed over a shorter distance. Those would be harder to discern in the \HI map. We may also underestimate the incident UV flux, although not by much more than the GALEX detection limit in this case, so this effect would be small, but may be amplified if the separation between the UV source and the \HI patch is small.

Only near position 2 do we find, perhaps because of a favourable morphology of the gas, an HI patch sufficiently close to the UV source to trace a higher-density GMC. This position was not included in the candidate PDRs selected by \citet{Heiner2008a}, who did not aim to study a complete set of FUV regions in M81. In contrast, we do not detect any of the GMCs traced with the PDR method in their paper, potentially because the low CO emission in those regions and M81 overall is insufficient to be traced in our map.

%
%
%
%

\section{Line ratios}
\label{sec:LineRatiosGasMass}

We calculated the CO $(3-2)/(1-0)$ line ratios for each of our 9 CO $J=3-2$ detections using the CO $J=1-0$ measurements from \citet{Brouillet1988} derived from NRAO observations. To compare both data sets, we convolved our CO $J=3-2$ cube to match the resolution of the Brouillet data (60 arcsec; see Section \ref{sec:COBrouillet}). A Gaussian fit was then made to each line. In Table \ref{tab:LineRatios}, we show the values and ratios for the line  fluxes. At 60\,arcsec resolution, positions 2 and 9 cannot be separated from 1 and 6, respectively, thus we give the average values. The Brouillet map does not cover position 8.

The CO $(3-2)/(1-0)$ ratios for the integrated flux of each peak lie in the range $(0.2-0.8)$ with a mean value of $0.43\pm0.18$, which is consistent with previous results from the JCMT NGLS (\citealt{Wilson2009,Warren2010,Bendo2010,Irwin2010}, B.~Tan et al. in prep.). 

Using the JCMT data of NGC~4631 and a single component model, \citet{Irwin2010} calculate the relation between the $J=(3-2)/(1-0)$ ratio and the hydrogen density for a number of temperatures. Assuming temperatures from 10$\,\rm K$ to 50$\,\rm K$ and a CO $(3-2)/(1-0)$ ratio in the range 0.2 to 0.8, this model constrains the hydrogen density, $n\rm(H_2)$, in M81 to a range between $10^3$ and $10^4\,\rm cm^{-3}$.

\citet{Israel2009} proposed the use of a two component model as the best approach to studying the molecular gas dynamics. In this model, the warm ($T\rm_{kin}\sim150\,K$) gas represents just a small part of the overall molecular gas content, which is mostly ($\sim80\%$) cold ($T\rm_{kin}\sim10\,K$). With this additional constraint, the hydrogen density for M81 is most likely to be near the lower end of the range indicated above.

\begin{table}
   \caption{Line ratios.}
   \label{tab:LineRatios}
   \centering
   \begin{tabular}{cccc}
       \hline
          & CO $J=3-2$ flux & CO $J=1-0$ flux & (3-2)/(1-0)\\
          & $\rm K\,km\,s^{-1}$ & $\rm K\,km\,s^{-1}$ & \\
       \hline
       1    &  0.33  &  0.8  &  0.41   \\
       2    &  0.24  &  0.8  &  0.31   \\ 
       3    &  0.41  &  0.5  &  0.82   \\
       4    &  0.22  &  0.8  &  0.28   \\
       5    &  0.15  &  0.7  &  0.21   \\
       6    &  0.22  &  0.5  &  0.44   \\
       7    &  0.26  &  0.5  &  0.52   \\
       8    &  0.15  &   -   &   -     \\
       9    &  0.31  &  0.7  &  0.44   \\ \hline
       Mean &    -   &    -  &  0.43   \\
       \hline
   \end{tabular}
\end{table}

%
%
%
%

\section{Concluding remarks}
\label{sec:Conclusions}

We have presented a new CO $J=3-2$ map of M81 obtained as part of the JCMT NGLS, as well as a new H$\alpha$ image from the INT, and combined these new data with public \HI, IR, and FUV data. The study of these data leads us to the following conclusions:

\begin{itemize}

\item[(i)] We confirm an overall low level of CO emission in M81. The emission that we do detect is limited to a few small regions within $D_{25}/2$, all of which are located in the spiral arm region. These results agree qualitatively with those previously published for the $J=1-0$ transition of CO. Moreover, the regions where the CO is detected also seem to be the areas where the star formation is more conspicuous (see Figure \ref{fig:Maps}).

\item[(ii)] We compare the spectral profiles of our CO $J=3-2$ detections with those of \HI data at the same positions. There is good agreement between the peak velocities and line widths in both cases, which confirms our CO detections and indicates that the kinematical conditions of the molecular and atomic gas are similar.

\item[(iii)] We confirm the previous results by \citet{Allen1997} that, locally, the H$\alpha$ emission is almost without exception well traced by the FUV emission. The opposite is not always true, and some features in the FUV map do not have any corresponding detection in H$\alpha$. The same holds when comparing H$\alpha$ with the 24$\,\mu$m emission.

\item[(iv)] We found statistical evidence supporting the relation between molecular gas and star formation in M81, although the correlation between the star formation tracers is much stronger.

\item[(v)] We investigated candidate PDRs close to the CO $ J=3-2$ detections and find total hydrogen volume densities in the range $1-178\,\rm cm^{-3}$, in agreement with other candidate-PDRs studied in \citet{Heiner2008a}. The morphology at the location of the highest density is indicative of a relatively small, compact PDR, as opposed to the other locations where such a morphology could not be seen.

\item[(vi)] We calculated $J=(3-2)/(1-0)$ line ratios for each of our detections, obtaining values in the range $(0.2-0.8)$ with a mean value of 0.43$\pm0.18$. This is consistent with previous work for other galaxies in the NGLS. Using a single component model we constrained the hydrogen density to the range $(10^3-10^4)\,\rm cm^{-3}$, most likely near the lower value of this interval. 

\end{itemize}

\section*{Acknowledgments}

We thank Leonel Guti\'errez for providing several scripts that were extremely useful for our work. We also thank Mike Irwin and Eduardo Gonz\'alez for applying the INT reduction pipeline to our images, making the data analysis process remarkably easy. We are deeply grateful to Elias Brinks for his very useful comments on a previous version of this manuscript. Finally, we thank the anonymous referee for her/his thorough review which significantly contributed to improve the quality of this work.

The James Clerk Maxwell Telescope is operated by The Joint Astronomy Centre on behalf of the Science and Technology Facilities Council of the United Kingdom, the Netherlands Organisation for Scientific Research, and the National Research Council of Canada. Based on observations made with the Isaac Newton Telescope operated on the island of La Palma by the Newton Group of Telescopes in the Spanish Observatorio del Roque de los Muchachos of the Instituto de Astrof\'{i}sica de Canarias. This work is based in part on observations made with the Spitzer Space Telescope, which is operated by the Jet Propulsion Laboratory, California Institute of Technology under a contract with NASA. Support for this work was provided by NASA. This research has made use of the NASA/IPAC Extragalactic Database (NED) which is operated by the Jet Propulsion Laboratory, California Institute of Technology, under contract with the National Aeronautics and Space Administration. We acknowledge the use of data from THINGS, 'The \HI Nearby Galaxy Survey' \citep{Walter2008}. JSH is grateful for a postdoctoral fellowship from the Centre de Recherche en Astrophysique du Qu\'{e}bec (CRAQ).\\

\bibliographystyle{aa}
\bibliography{Papers}
\nocite{Knapen2008}

%
%
%
%

\end{document}